 \documentstyle[12pt]{article}

 \let\a=\alpha \let\b=\beta \let\g=\gamma \let\d=\delta \let\e=\epsilon
  \let\q=\theta  \let\k=\kappa
\let\l=\lambda \let\m=\mu \let\n=\nu   \let\r=\rho
\let\s=\sigma   \let\f=\phi  \let\y=\psi
      \let\G=\Gamma \let\D=\Delta \let\Q=\Theta \let\L=\Lambda
 \let\P=\Pi   \let\F=\Phi \let\Y=\Psi
 
\let\la=\label \let\ci=\cite 
  
\def\nn{\nonumber} \def\bd{\begin{document}} \def\ed{\end{document}}
\def\ds{\documentstyle} \let\fr=\frac \let\bl=\bigl \let\br=\bigr
\let\Br=\Bigr \let\Bl=\Bigl
\let\bm=\bibitem
\let\na=\nabla
\let\pa=\partial \let\ov=\overline
\newcommand{\be}{\begin{equation}}
\newcommand{\ee}{\end{equation}}
\def\ba{\begin{array}}
\def\ea{\end{array}}
\newcommand{\ho}[1]{$\, ^{#1}$}
\newcommand{\hoch}[1]{$\, ^{#1}$}
\newcommand{\bea}{\begin{eqnarray}}
\newcommand{\eea}{\end{eqnarray}}
\newcommand{\ra}{\rightarrow}
\newcommand{\lra}{\longrightarrow}
\newcommand{\Lra}{\Leftrightarrow}
\newcommand{\ap}{\alpha^\prime}
\newcommand{\bp}{\beta^\prime}
\newcommand{\tr}{{\rm tr} }
\newcommand{\Tr}{{\rm Tr} }
\newcommand{\NP}{Nucl. Phys. }
\newcommand{\tamphys}{\it Center for Theoretical Physics\\
Physics Department \\ Texas A \& M University
\\ College Station, Texas 77843}

\begin{document}

\hfill{CTP-TAMU-20/93 }

\vspace{24pt}

\begin{center}

{ \large {\bf  Higher Spin BRS Cohomology of  Supersymmetric
  Chiral Matter in D=4 }}

\vspace{36pt}

J. A. Dixon, R. Minasian and J. Rahmfeld

\vspace{6pt}

{\tamphys}

\vspace{6pt}

\vspace{6pt}

\underline{ABSTRACT}

\end{center}
We examine the BRS cohomology of chiral matter in $N=1$, $D=4$ supersymmetry to
determine a general form of composite superfield operators which can suffer
from supersymmetry anomalies. Composite superfield operators $\Y_{(a,b)}$ are
products of the elementary chiral superfields $S$ and $\ov S$ and the
derivative operators $D_\a$, $\ov D_{\dot \b}$ and $\pa_{\a \dot \b}$. Such
superfields $\Y_{(a,b)}$ can be chosen to have `$a$' symmetrized undotted
indices $\a_i$ and `$b$' symmetrized dotted indices $\dot \b_j$. The result
derived here is that each composite superfield $\Y_{(a,b)}$ is subject to
potential supersymmetry anomalies if $a-b$ is an odd number, which means that
$\Y_{(a,b)}$ is a fermionic superfield.

\vfill

\baselineskip=12pt

\pagebreak

\setcounter{page}{1}

\hsize=6.5truein
\textheight=8.5truein
\topmargin=.1in
\evensidemargin=0in
\oddsidemargin=0in
\headsep=0in
\section{Introduction}
\setcounter{equation}{0}
\renewcommand{\theequation}{1.\arabic{equation}}

The only known candidate for  a unified theory of all matter and forces
is superstring theory, but there are two major obstacles to making a
comparison between this theory and experiment.  The first problem is to
discover how and why supersymmetry gets broken, preferably without
generating a ridiculously huge cosmological constant.  The second problem
is to explain why our own universe is picked out from other possibilities.
In a recent  book written for the general public \ci{weinberg}, Weinberg
 has expressed some doubt whether either
of these questions has a mathematical answer--and suggested that the
explanation may simply be that if our universe were not as it is,
we wouldn't be here to ask the question.

But of course this `explanation' is a last resort.  Our purpose here is
to continue the search for supersymmetry anomalies.  If these exist,
their elimination would naturally be expected to
impose restrictions on the possible superstring theories.  In addition,
it has been conjectured \ci{erice} that such anomalies might also provide a
natural mechanism whereby `supersymmetry breaks itself', while at
the same time retaining the cosmological constant at the zero value
it naturally has in many unbroken supersymmetric theories.

The essential missing link in this program is that, as yet, there
has been no calculation of a non-zero
coefficient for any supersymmetry anomaly.  Efforts in this
direction will be reported elsewhere.

In this paper, we work out in detail the cohomology of the BRS
operator defined by the supersymmetry invariance of
chiral multiplets of rigid $N=1,D=4$ supersymmetry.
The new result here is that this cohomology space
contains potential anomalies in the renormalization of
fermionic superfields with all half-integer spins.
Formerly it had been shown that there were potential anomalies
for fermionic superfields with spin $\fr{1}{2}$ only.

This may be very important for superstring theories, since such
higher spin multiplets necessarily occur in all such theories.

\section{Summary of Previous Work}
\setcounter{equation}{0}
\renewcommand{\theequation}{2.\arabic{equation}}

A systematic method for  the calculation of local BRS cohomology spaces was
described in \cite{cmp1}.  The method starts with the definition
of a grading operator which
generates a  `spectral' sequence of simpler nilpotent operators
whose cohomology spaces are easily found. This sequence of spaces converges to
a space isomorphic to the desired cohomology space.  To facilitate computation
of each of the cohomology spaces, we introduce a Fock space so that each
successive cohomology space is the kernel of a `Laplacian'  operator.

The cohomology space enables one to determine whether the theory can possess
anomalies, either in the renormalization of the action itself or in the
renormalization of higher dimensional composite operators formed by the fields
in the theory.
In most cases, it is quite arduous to analyze the cohomology space of a field
theory in this general way, especially when the space itself is nontrivial to
describe, as  is frequently the case. The cohomology of
Yang-Mills theory was examined in a specific case in \cite{cmp1}.
An investigation was done
 of the simplest supersymmetric theory in four dimensions,
the Wess-Zumino chiral theory, in \cite{cmp2}.
The present work completes those results.
The results of \cite{cmp2} were generalized in \ci{d1} and \ci{prl}  to include
the case where chiral matter is coupled
to supersymmetric Yang-Mills theory.
 Most recently, this method was used in a general study of the
cohomology of the supertranslation operator \ci{dm}.
The results of \ci{dm} and the results here are closely related, since in both
cases the cohomology is determined by Laplacians which involve only counting
operators and coupled $SU(2)$ angular momentum operators.

A different approach was used by the authors of \ci{bdk1}, where a general
formula for the creation of Lorentz invariant polynomials in the cohomology
space for all compact gauge groups for the restricted BRS operator in
Yang-Mills theories and gravity is given. Some aspects of the BRS cohomology of
supersymmetric theories in four dimensions, restricted to Lorentz invariant
polynomials, are investigated in \ci{b1,b2}. The cohomology of local integrated
polynomials in field theories was also examined recently in \ci{dhtv1,dhtv2}.
Spectral sequences have also been used in the
BRS  cohomology of $2d$ gravity (but
without the introduction of a positive Fock space metric)
\cite{lian}; for a review of this technique applied to CFT and $2d$ gravity
see e. g. \cite{bmp}.

\section{Action and Supersymmetry Invariance for Wess-Zumino Model}
\setcounter{equation}{0}
\renewcommand{\theequation}{3.\arabic{equation}}

 The first attempt  at the calculation of the BRS cohomology of the Wess-Zumino
theory was made in \cite{cmp2}. That paper used Majorana spinors and real four
dimensional Dirac matrices $\g_{\m}$. It is easier to obtain those results and
also possible to construct the complete solution by using complex two component
Weyl spinors and the matrices $\s^{\m}$.

One advantage of the complex two component notation\footnote
{  Our notation and some useful identities are discussed in
\ci{dm}. } is that the superfield
formalism is much easier when expressed in terms of Weyl spinors  since  the
chiral constraint is essentially complex in nature and the theory is symmetric
under complex conjugation.  In this notation,
it becomes clear that the very complicated operators
of \ci{cmp2} are really coupled angular momenta as in
\ci{dm}.

In complex two-component notation, the (free quadratic) action for the
Wess-Zumino chiral model is:
\begin{equation} S = -
\int d^{4}x [\partial_{\mu} A \partial^{\mu} {\ov A}
+ \y^{\a} \sigma^{\m}_{\;\a \dot \b} \partial_{\mu} \ov{\y}^{\dot \b}
- F \ov{ F}]. \la{action}
\ee
This action is invariant under the following supersymmetry and
translational transformations (which are assumed to imply their complex
conjugates):
\begin{equation} \delta A = c^{\a} \y_{\a}
+ \epsilon^{\mu} \partial_{\mu} A \end{equation}
\begin{equation}
\delta \y_{\a} = \pa_{\m} A
\s^{\m}_{\a \dot{\b}}
\overline{c}^{\dot{\b}}
+ F c_{\a}
+ \epsilon^{\mu} \partial_{\mu} \y_{\a}
\ee
\be
\delta F = \pa_{\m} \y^{\a} \s^{\m}_{\a \dot{\b}} \overline{c}^{\dot{\b}}
+ \epsilon^{\mu} \partial_{\mu} F.
\ee
Here $c^{\a}$ is a constant (spacetime independent), commuting
two component complex chiral
spinor and $\epsilon_{\mu}$ is a constant real anticommuting Lorentz
vector.  Their variations are:
\begin{equation} \delta
\epsilon^{\mu} = - c^{\a}  \s^{\m}_{\a \dot{\b}} \overline{c}^{\dot{\b}}
= - c \cdot  \s^{\m} \cdot \overline{c}
\ee
\begin{equation} \delta c^{\a} = 0. \end{equation}
It is straightforward to show that
\begin{equation} \delta^{2} = 0  \end{equation}
on any field (including $\epsilon_{\mu}$ and $c^{\a}$ as constant fields).
Note that
\be
(c^{\a}  \s^{\m}_{\a \dot{\b}} \overline{c}^{\dot{\b}} )^*
=
c^{\b}  \s^{\m}_{\b \dot{\a}} \overline{c}^{\dot{\a}}
\ee
is a real quantity.

Another way to express the generator $\d$ is
\[
\d = \int d^4 x \Bl \{
[c^{\a} \y_{\a}
+ \epsilon^{\mu} \partial_{\mu} A] \fr{\d}{\d A}
+[
\pa_{\m} A
\s^{\m}_{\a \dot{\b}}
\overline{c}^{\dot{\b}}
+ F c_{\a}
+ \epsilon^{\mu} \partial_{\mu} \y_{\a}]
\fr{\d}{\d \y_{\a}}
\]
\[ +
[
\pa_{\m} \y^{\a}
\s^{\m}_{\a \dot{\b}}
\overline{c}^{\dot{\b}}
+  \epsilon^{\mu} \partial_{\mu} F]
\fr{\d}{\d F}
+ [ {\ov c}^{\dot{\a}} \ov{\y}_{{\dot \a}}
+ \epsilon^{\mu} \partial_{\mu} {\ov A}] \fr{\d}{\d {\ov{A}}}
\]
\[
+[
\pa_{\m} {\ov A}
{\ov \s}^{\m}_{{\dot \a} \b}
{c}^{\b}
+ {\ov F}{\ov c}_{{\dot \a}}
+ \epsilon^{\mu} \partial_{\mu} \ov{\y}_{\dot{\a}}]
\fr{\d}{\d \ov{\y}_{\dot \a} }
\]
\be + [
\pa_{\m} {\ov \y}^{{\dot \a}}
\ov{\s}^{\m}_{\dot{ \a} \b }
c^{\b}
+  \epsilon^{\mu} \partial_{\mu} {\ov F}]
\fr{\d}{\d{\ov F}}
\Br \}
-  c^{\a} \s^{\m}_{ \a \dot{\b}} {\ov c}^{\dot{\b}}
\fr{\pa}{\pa \epsilon^{\mu}}.
\la{delta}
\ee

\section{The Grading of the Spectral Sequence}
\setcounter{equation}{0}
\renewcommand{\theequation}{4.\arabic{equation}}

The final goal is to find the cohomology space
\be
H \approx \mbox{Ker}\, \d / \mbox{Im}\, \d ,
\ee
which is isomorphic to the kernel of the Laplacian
\be \Delta =\left(\delta +\delta^{\dag} \right)^{2}.
\ee
Unfortunately, $\D$ is in general a very complicated operator, and it is not
possible to deduce much about ${\rm kernel}\; \D$
 from the expression for $\D$  directly--one just gets a huge
number of terms and no insight.

The key idea behind the spectral sequence
formalism\footnote{A complete
discussion of the spectral sequence method for finding
BRS cohomology can be found in \cite{cmp1}.}
is to divide
$\d$ into parts which are easier to work with. For this purpose, we need to
define a suitable counting operator $N_{\rm grading}$,
which assigns  to each term of $\d$
a positive (or zero) integral order.
We decompose
\be \d =\sum_{i=0}^{\infty}\d_i ,
\ee
with
\be
[N_{\rm grading}, \d_i] = i \d_i \label{grading}.
\ee
The grading is certainly not uniquely defined, and an important and difficult
part of the  spectral sequence technique is to find a useful grading.
The spectral sequence generated by a given grading consists of a sequence of
positive semidefinite Laplacian operators
\be
 \D_r =\left( d_r + d_r^{\dag} \right)^{2}, \, r\geq 0 \ee
where each successive nilpotent operator $d_{r+1}$ operates in the
cohomology space $E_{r+1}$, defined by $E_{r+1}=\mbox{ker} \, \D_{r}$.
The spaces satisfy the relation ($E_0$ is the whole space in which $\d$ acts):
\be
E_{\infty} \subseteq ... \subseteq E_{r+1} \subseteq E_r \subseteq ...\subseteq
   E_0.
\ee
In practice, the sequence collapses (i.e. $d_r=0$ for $r\geq r_0$
so that $E_{\infty}=E_{r_0}$) for some low value  $r_0$ of $r$
($r_0=3$ in the
present case).

For the present problem, we will use
the counting operator
\be
 N_{\rm grading}= N_{\rm grad}
+ {\ov N}_{\rm grad}
\la{ngrad}
\ee
with
\be
N_{\rm grad}
= 3 N(A) + 2 N(\y) + N(F) + N(c).
\la{gradop}
\ee
We easily see that the decomposition
\be
\d = \d_0 + \d_2 \label{delpart}
\ee
fulfills (\ref{grading}), where
\be
 \d_0
 = c^{\a} \L_{\a}
+
\ov{c}^{\dot \a} \ov{\L}_{\dot \a}
+
\e^{\m} \pa_{\m} \label{delta0}
\ee
and
\be
 \d_2  = c^{\a} {\ov \na}_{\a}
+ \ov{c}^{\dot \a} \na_{\dot \a}
-  c^{\a} \s^{\m}_{ \a \dot{\b}} {\ov c}^{\dot{\b}}
(\epsilon^{\mu})^{\dag} . \label{delta2}
\ee
Here we define:
\be
\L_{\a} =
\int d^4 x \Bl \{
\y_{\a}
\fr{\d}{\d A}
+ F
\fr{\d}{\d \y^{\a}}
\Br \} \label{lambda}
\ee
\be
\ov{\L}_{\dot \a} =
\int d^4 x \Bl \{
{\ov \y}_{\dot{\a}}
\fr{\d}{\d \ov{A}}
+{\ov F}
\fr{\d}{\d \ov{\y}^{\dot{\a}}}
\Br \} \label{lambdadag}
\ee
\be
{\ov \na}_{ \a}
=\int d^4 x \Bl \{
\pa_{\m} {\ov A}
\s^{\m}_{\a \dot{\a}}
\fr{\d}{\d \ov{\y}_{\dot \a}}
+
\pa_{\m} \ov{\y}^{\dot \a}
\s^{\m}_{\a \dot{\a}}
\fr{\d}{\d \ov{F} }
\Br \}
\ee
\be
\na_{ \dot \a}
=\int d^4 x \Bl \{
\pa_{\m}  A
\s^{\m}_{\a \dot{\a}}
\fr{\d}{\d \y_{\a}}
+
\pa_{\m} \y^{ \a}
\s^{\m}_{\a \dot{\a}}
\fr{\d}{\d F }
\Br \}.
\ee
The grading (\ref{ngrad}) separates $\d$ into  parts with no derivatives ($\L$)
and with derivatives $\left( \na , \, \pa \right)$ and is
appropriate for this problem, because
the $\L$ part is particularly easy to deal
with. $\pa_{\m}$ was already fully analyzed in \ci{cmp1}.

\section{The Operator $\D_0$}
\setcounter{equation}{0}
\renewcommand{\theequation}{5.\arabic{equation}}

Now that we have chosen a grading, we go through the steps of the
spectral sequence. The starting point is to calculate the kernel of
the operator $\D_0 = [\d_0 + \d_0^{\dag}]^2$.
{}From (\ref{lambda}) and (\ref{lambdadag}), it follows that:
\be
\{ \L_{\a}, \L_{\b} \}=0
\ee
\be
\{ \L_{\a}, (\L_{\b})^{\dag} \}
=
\d_{\a}^{\b} N \label{antila}
\ee
\be
\{ \L_{\a}, (\ov{\L}_{\dot \b})^{\dag} \}
=
0
\ee
and, therefore, $\D_0$ computed from (\ref{delta0}) takes the form:
\[
\D_0 =
(\L_{\a})^{\dag}
\L_{\a} +
(\ov{\L}_{\dot \a})^{\dag}
\ov{\L}_{\dot \a}
+ \pa_{\m}
( \pa_{\m})^{\dag} +(N+{\bar N})\e_{\m}^{\dag}\e_{\m}
\]
\be
+ n N
+ \ov{n} \ov{N}.
\label{del0}
\ee
In (\ref{del0}) we used the counting operators
\be
n  = c_{\a} ( c_{\a})^{\dag}
\ee
\be
\ov{n} = \ov{c}_{\dot \a} ( \ov{c}_{\dot \a})^{\dag}
\ee
and
\be
N = \int d^4 x \; \Bl \{ A \fr{\d}{\d A}
+ \y_{\a} \fr{\d}{\d \y_{\a}}
+ F  \fr{\d}{\d F} \Br \},
\ee
which
can also be written as
\[ N
=
\sum_{k=0}^{\infty} \fr{1}{k!}
\Bl \{
 A_{\m_1 \m_2 \ldots \m_k}
(A_{\m_1 \m_2 \ldots \m_k})^{\dag}
\]
\be
+\y_{\a \m_1 \m_2 \ldots \m_k}
(\y_{\a \m_1 \m_2 \ldots \m_k})^{\dag}
+F_{\m_1 \m_2 \ldots \m_k}
(F_{\m_1 \m_2 \ldots \m_k})^{\dag}
\Br \}.
\la{number}
\ee
Here the definition
\be
A_{\m_1 \m_2 \ldots \m_k}
=
\pa_{\m_1} \pa_{\m_2} \ldots \pa_{\m_k} A
\ee
is used (same for $F$ and $\y_{\a}$).
(\ref{del0}) also uses
the relation:
\be
[\pa_{\m}^{\dag},\pa_{\n}]=\d^{\m}_{\; \n} (N+{\ov N}) \label{dddag}
\ee
which was discussed in \ci{cmp1}.

\section{The Space $E_1$}
\setcounter{equation}{0}
\renewcommand{\theequation}{6.\arabic{equation}}
Since (\ref{del0}) consists of a sum of separately positive semidefinite
operators in the
form \newline  $\sum_{i}~Q_{i}^{\dag}~Q_i~=~\D$,
the kernel satisfies the equations
\be
Q_i \; \mbox{ker}\, \D=0,
\ee
or,
more specifically:
\be
\L_{\a} E_1 = 0
\la{lam}
\ee
\be
\ov{\L}_{\dot \a} E_1 =0
\la{lambar}
\ee
\be
( \pa_{\m})^{\dag} E_1 = 0
\la{par}
\ee
\be
n N E_1 = 0 \ee
\be
\ov{n} \, \ov{N} E_1 = 0
\la{lam5}
\ee
\be
\left( N+ {\bar N} \right) \e_{\mu} E_{1}=0 \label{lam6}
\la{nnn}.
\ee
The solutions of these constraints are much
more obvious than they were in the real notation of \ci{cmp2}.
If $nN=0$ then either $n=0$ or $N=0$. Hence,
any function of the form
\be
{\cal F}=
{\cal F}(\pa, A, F, \y, \ov{ c}) \Q
\la{func1}
\ee
satisfies this whole set of equations, if it fulfills (\ref{lam})
and (\ref{par}). Functions of this kind and their complex conjugates exhaust
the solutions of these equations depending non-trivially on $c$ or $\ov c$.

Solutions independent of $c$ and $\ov c$ can
depend on all variables
\be
{\cal F}
=
{\cal F}
(\pa, A, F, \y, \ov{A},
\ov{ F},\ov{ \y}) \Q
\la{func2}
\ee
and must satisfy all the equations (\ref{lam}), (\ref{lambar}) and
(\ref{par}).

The expression $\Q$ is defined as
\be
\Q = \e^{\m}  \e^{\n}  \e^{\l}  \e^{\r} \varepsilon_{\m \n \l \r}
\ee
and satisfies
\be
\e_{\m}^{\dag} \e_{\m} \Q =
( 4 - N(\e) ) \Q = 0
\la{eps}
\ee
We can think of $\Q$ as being equivalent to $\int d^4 \, x$ (see \cite{cmp1}).

To construct  the operator $d_1$ we will need the explicit form of
the operator $\P_{1}$,
which projects the entire space $E_0$ onto
$E_{1}$. A general form
is easy to write down:
\bea
\P_1 & = &
\P_{ \e = 0 }
\P_{ \pa^{\dag} = 0 }
\Bl \{
\P_{ N \geq 0} \P_{ \ov{N} \geq 0 }
\P_{\L = 0}\P_{\ov{\L} = 0} \P_{n=0}\P_{\ov{n} = 0} +\nonumber \\
& &
+
\P_{ N > 0 }
\P_{ n =  0 }
\P_{ \ov{N} = 0 }
\P_{ \ov{n} > 0 }
\P_{ \L = 0}
+\P_{ \ov{N} > 0 }
\P_{ \ov{n} =  0 }
\P_{ N = 0 }
\P_{ n > 0 }
\P_{ \ov{\L} = 0}
\Br \}  \nonumber \\
& & + \P_{ N =0}\P_{\ov{N} = 0 }.
\la{project}
\eea
Here, the first part projects onto solutions of type (\ref{func2}).
The second operator projects onto  states of  type (\ref{func1}) and the
third projects onto their complex conjugates, i.e.
\be
 {\cal F}={\cal F} (\pa, \ov A, \ov F, \ov \y, { c}) \Q .
\ee
 The fourth operator
projects onto pure ghost states, formed by
 $\epsilon_{\m}, {c}^{\a} $ and
$ {\ov c}^{\dot \a } $ only. One can easily verify that these states satisfy
(\ref{lam}) - (\ref{nnn}) also, but they do not contain any information
of interest for this paper (see \cite{dm} for a discussion of these terms).

Now we find explicit forms of the operators contained in (\ref{project}).
In addition to the defining equations for an
orthogonal projection operator
$\P^2 =\P =\P^{\dag}$, $\P_{ \L=0 }$ is subject to the constraint
\be
\L_{\a} \P_{ \L=0 }=0. \label{laproco}
\ee
It is easy to see that
the projection operator
\[
\P_{\L =0} =
1-\fr{1}{N} \L_{\a}^{\dag} \L_{\a}
+\fr{1}{2N^2} \L_{\a}^{\dag} \L_{\b}^{\dag} \L_{\b} \L_{\a}
\]
\be
=\fr{1}{4N^2 }
\L^{\a}
\L_{\a}
(\L^{\b}
\L_{\b})^{\dag} \label{lapro}
\ee
satisfies (\ref{laproco}). Note that the $\L_{\a}$ are anticommuting
objects, so that (\ref{lapro}) involves only quadratic terms ($\L^3=0$).

Similarly,
\be
\P_{\ov{ \L} =0} =\frac{1}{{4\ov N}^{2}}\ov{\L}^{\dot \a}
\ov{\L}_{\dot \a}
(\ov{\L}^{\dot \b}
\ov{\L}_{\dot \b})^{\dag} \label{ladapro}
\ee
fulfills the corresponding condition for $\ov{\L}$.
Equation (\ref{par}) requires an operator satisfying
\be
\pa_{\m}^{\dag} \P_{ \pa^{\dag}=0 }=0. \la{papro}
\ee
This projection operator has the form
\be
\P_{ \pa^{\dag}=0 }=
\sum_{k=0}^{\infty} \Bl \{ \fr{-1}{N+{\ov N}} \Br \}^k
\pa_{\m_1} \pa_{\m_2} \cdots \pa_{\m_k}
\pa_{\m_k}^{\dag} \cdots \pa_{\m_2}^{\dag} \pa_{\m_1}^{\dag}
\la{der}
\ee
as can be shown using the
commutation relation (\ref{dddag}).
The projection operators
$\P_{N=0}$ and $\P_{N>0}$
for the counting operators are
\be
 \P_{N=0 }=
\sum_{k=0}^{\infty}  \fr{(-1)^{k}}{k!}N_k
\la{N0}
\ee
with $N_k=\sum_{ \{ i\} }[\f_{i_{1}}\f_{i_{2}}.... \f_{i_{k}}]
                 [\f_{i_{1}}\f_{i_{2}}.... \f_{i_{k}}]^{\dag}$. $\f$
represents any relevant set of fields.
Of course, $\P_{N>0}$ is given by $\P_{N>0}=1-\P_{N=0}$.

\section{The Operator $\D_2$}
\setcounter{equation}{0}
\renewcommand{\theequation}{7.\arabic{equation}}
The operator $d_1=\P_1 \d_1 \P_1$ vanishes since $\d_{1}=0$, which results in
$E_{2}=E_{1}$ and $\P_2=\P_1$. Then
$d_2$ reduces to
\bea
d_2  & = &
\P_2 \Bl \{
c^{\a} {\ov \na}_{\a}
+
\ov{c}^{\dot \a} \na_{\dot \a}
-  c^{\a} \s^{\m}_{ \a \dot{\b}} {\ov c}^{\dot{\b}}
(\epsilon^{\mu})^{\dag} \Br \}
\P_2
 \nn \\
& = &
\P_2 \Bl \{
c^{\a} {\ov \na}_{\a}
+
\ov{c}^{\dot \a} \na_{\dot \a}
 \Br \}
\P_2 \P_{\e=0} - \P_{N=\ov N=0}
   \left(c^{\a} \s^{\m}_{ \a \dot{\b}} {\ov c}^{\dot{\b}} \right)
(\epsilon^{\mu})^{\dag}.
\eea
Now we concentrate on the sector where either $N\neq 0$ or $\ov N \neq 0$
or both. Then the operator
$c^{\a} \s^{\m}_{ \a \dot{\b}} {\ov c}^{\dot{\b}} (\e^{\mu})^{\dag}$ can be
dropped. Its cohomology can be found in \cite{dm}.
 From the  general theory of spectral
sequences, it follows that
$d_{2}$ is nilpotent, as can be verified explicitly.
To compute $\D_{2}$, we first evaluate some (anti)commutators.
It is a remarkable fact that these (anti)commutators frequently give
rise again to the operators we start with (or their adjoints):
\be
\{ {\ov \na}_{\a}, (\L_{\b})^{\dag} \} =0
\ee
\be
\{ {\ov \na}_{\a}, (\ov{\L}_{\b})^{\dag} \} =0
\ee
\be
[ \pa_{\m}, \L_{\a} ] =0 \la{lapa}
\ee
\be
[ \pa_{\m}, (\L_{\a})^{\dag} ] =0 \la{ladagpa}
\ee
\be
[ \pa_{\m}, {\ov \na}_{\a}] =0
\ee
\be
[ (\pa_{\m})^{\dag}, {\ov \na}_{\a} ] =
\s^{\m}_{\a \dot{\b} } \left(\ov{\L}_{\dot \b}\right)^{\dag}
\ee
\be
\{ \L_{\a}, \na_{\dot \b} \}
=
\s^{\m}_{\a \dot{\b} } \pa_{\m}
\ee
\be
\{ \L_{\a}, {\ov \na}_{ \b} \}
= 0
\ee
\be
\{ {\ov \na}_{\a}, ({\ov \na}^{\b})^{\dag} \}
=
\varepsilon_{\a \b} {\ov M} +
\s^{\m \n }_{\a \b } {\ov L}_{\m \n}.
\ee
Using the above relations, we find
\be
\P_2 {\ov \na}_{\a} \P_2 =
\P_2 {\ov \na}_{\a} \la{pinapi}
\ee
and hence,
\bea
\D_2 & =&
\P_2
\Bl \{
c^{\a} (c^{\b})^{\dag}
\{ {\ov \na}_{\a}, ({\ov \na}_{\b})^{\dag} \}
+
\ov{c}^{\dot \a} (\ov{c}^{\dot \b})^{\dag}
\{ \na_{\dot \a}, ( \na_{\dot \b})^{\dag} \}
 \nonumber \\
 & &
+ c^{\a} (c^{\b})^{\dag}
({\ov \na}_{\b})^{\dag} (\P_2 -1) {\ov \na}_{\a}
+
\ov{c}^{\dot \a} (\ov{c}^{\dot \b})^{\dag}
 ( \na_{\dot \b})^{\dag} (\P_2 -1) \na_{\dot \a}
\nonumber \\
& &
+
({\ov \na}_{\a})^{\dag} \P_2
{\ov \na}_{\a}
+
(\na_{\dot \a})^{\dag} \P_2
\na_{\dot \a}
\Br \}
\P_2
\nonumber \\
 &=&
\P_2
\Bl \{
({\na}_{\dot \a})^{\dag} \P_2 {\na}_{\dot \b}
+
{\ov n}  [ M -4 ]
-
4 {\ov J}_i L_i
\nonumber \\
& &
+
({\ov \na}_{\a})^{\dag} \P_2 {\ov \na}_{\a}
+
{n}  [ {\ov M} -4 ]
-
4 {J}_i {\ov L}_i
\Br \}
\P_2 . \label{del2}
\eea
We use the following abbreviations
\be
M = N({ \pa}) + 4 N(F)  + 2 N(\y)
\ee
\be
{\ov M} = {\ov N}(\pa) + 4 N({\ov F})  + 2 N({\ov \y})
\ee
\be
N(\pa) = \sum_{k=0}^{\infty}
\fr{1}{k!}
[ \f_{\m_1 \cdots \m_{k +1}  }
(\f_{\m_1 \cdots \m_{k+1}} )^{\dag}]; \;\;
\ee
\be
{\ov N}(\pa) = \sum_{k=0}^{\infty}
\fr{1}{k!}
[ {\ov \f}_{\m_1 \cdots \m_{k +1}  }
({\ov \f}_{\m_1 \cdots \m_{k+1}} )^{\dag}] .
\ee
$N(\pa)$ and its complex conjugate
count the number of derivatives in each expression.

The coupled angular
momenta $\ov J_i L_i$ arise as follows. Computation yields the term
\be
\frac{1}{2} \ov c^{\dot \a} (\ov c^{\dot \b})^{\dag}
\left(\ov \s^{\m \n}\right)_{\dot \a \dot \b} L_{\m \n} \la{coupling}
\ee
where
\be
 L_{\m \n} = \sum_{k=0}^{\infty}
\fr{1}{k!}
[ \f_{\m_1 \cdots \m_k \m }
(\f_{\;\;\;\m_1 \cdots \m_k }^{\n })^{\dag}
-
\f_{\m_1 \cdots \m_k \n }
(\f_{\;\;\; \m_1 \cdots \m_k }^{ \m })^{\dag}]; \; \; \left(\f = A,\y_{\a},F
\right).
\ee
Now using the identities
\be
 \left(\ov \s^{0 i}\right)_{\dot \a \dot \b}=
       \left( \s^{i}\right)_{\dot \a \dot \b} =
i\; \varepsilon^{ijk}  \left( \s^{k}\right)_{\dot \a \dot \b},
\ee
(\ref{coupling}) can be written as $\ov J_i L_i$, where
\bea
{\ov J}_i & = & \fr{1}{2} {\ov c}^{\dot \a} ({\ov \s}_i)_{\dot \a}^{\;\;
\dot \b} ({\ov c}^{\dot \b})^{\dag} \\
L_i & = & - \fr{1}{2} L_{0i} -
\fr{i}{4} \e_{ijk} L_{jk}.
\eea
Equivalently, one could write
\be
 L_i\left(\s^i\right)_{\dot \a \dot \b}= L_{\m \n}
                   \left(\s^{\m \n}\right)_{\dot \a \dot \b}.
\la{noname}
\ee
It is easy to verify that $L_i$ and $J_i$ obey the
commutation rules of the $SU(2)$ Lie algebra:
\be
[J_i, J_j] = i \e^{ijk} J_k
\ee
\be
[L_i, L_j] = i \e^{ijk} L_k.
\ee
The complex conjugate equations are:
\be
[- {\ov J}_i, - {\ov J}_j] =   i \e^{ijk} (- {\ov J}_k)
\ee \be
[- {\ov L}_i, - {\ov L}_j]  =   i \e^{ijk} (- {\ov L}_k)
\ee
and we note that
\be
[J_i, L_j] = 0.
\ee
{}From this we see that the effect of the $L_i$ is to `rotate' the dotted
indices
that arise in derivatives $\pa_{\a \dot \b}$.
It follows that the Laplacian contains
 only counting operators, coupled angular
momenta and the operator $\P_2\na_{\dot \a}\P_2$.

\section{The space $E_3$}
\setcounter{equation}{0}
\renewcommand{\theequation}{8.\arabic{equation}}
Given the manifest $SU(2)$ structure of the Laplacian
\[
\D_2
=
\P_2
\Bl \{
({\na}_{\dot \a})^{\dag} \P_2 {\na}_{\dot \a}
+
{\ov n}  [ M -4 ]
-
2[( {\ov J}_i + L_i)
( {\ov J}_i + L_i)
- {\ov J}_i
 {\ov J}_i
-
L_i
L_i]
\]
\be
+
({\ov \na}_{\a})^{\dag} \P_2 {\ov \na}_{\a}
+
{n}  [ {\ov M} -4 ]
-
2[( J_i + {\ov L}_i)
( J_i + {\ov L}_i)
-
 J_i J_i
-
{\ov L}_i
{\ov L}_i ]
\Br \}
\P_2 ,
\label{del2,2}
\ee
the problem of finding the kernel
reduces to the determination of the eigenvalues and eigenstates
of the operators.
We assume $n=0, \; \ov n \neq 0$ here (if $n=\ov n = 0 $, we just get
(\ref{E31}) below.)

Now the eigenvalue of the $SU(2)$ operator $\ov J^{2}=\ov J_i\ov J_i$
is given by
(see \ci{dm})
\be
\ov J^2=\ov j(\ov j+1) \; \mbox{with}
 \; \ov j = \fr{\ov n}{2},
\ee
since the operator $\ov J^{2}$ is acting on polynomials in $\ov c_{\dot \a}$
which are totally symmetric.

As is clear from (\ref{noname}),  the operator $L_i$ rotates the dotted
indices occurring in derivatives. Hence, the
maximum possible eigenvalue of
$L^{2}=L_iL_i$ is $\frac{N(\pa)}{2}$. Therefore, we find
\be L^{2} =l(l+1) \; \mbox{with} \; l=\frac{N(\pa)-2s}{2}, \, s=0, 1, 2...
\left [\frac{N(\pa)}{2} \right],
\ee
where $\left [x\right]$ is the greatest integer in $x$.

The coupling operator of the two angular momenta
 leads to the term
\be
(\ov J+L)^2=k(k+1), \, \mbox{where} \;
k= \ov j+l-r,
\ee
and $r=0,1,2,...,
\mbox{min}\bl({\ov n},N(\pa)-2s\br)$.

Now it follows from (\ref{del2,2}) that
\bea
\D_2 &
= &
\P_2
\Bl \{
({\na}_{\dot \a})^{\dag} \P_2 {\na}_{\dot \a}
+
{\ov n}  [ 4 N(F) + 2 N(\y) - 4 +2r+2s] +2r[N(\pa)+1-r-2s]
\Br \}
\P_2 \nonumber \\
 & & +\mbox{complex conjugate}. \label{del2b}
\eea
The equation (\ref{lapro}) implies that $N(F)\geq 1$ or $N(\y)\geq 2$ for
each term in every polynomial in ker $\L_{\a}$ and since
$r\leq N(\pa)-2s$, the following terms in (\ref {del2b}) are positive
semidefinite:
\be
\P_2 {\na}_{\dot \b} E_3 = 0 \la{E31}
 \ee
\be
{\ov n}  [ 4 N(F) + 2 N(\y) - 4 +2r+2s] E_3=0
\la{E31b}
\ee
\be r[N(\pa)+1-r-2s]
E_3 = 0 \la{E32}.
\ee
The complex conjugates of these equations are also true of course, for the case
when $\ov n=0$ and $n\neq 0$.

The only possible solutions for (\ref{E32}) occur when $r=0$, hence,
the only
possible solutions of (\ref{E31b}) occur for $s=0$.
This result implies $k=\ov j+l$.
It follows from  this and $l=\frac{N(\pa)}{2}$, that the
polynomials in $E_{\infty}$ are totally symmetric in the dotted indices.

With $r=s=0$, (\ref{E31b}) reduces to
\be
{\ov n}  [ 4 N(F) + 2 N(\y) - 4]
E_3 = 0 \la{E33}.
\ee

\section{The Cohomology Spaces
$E_{\infty}$ and $H$}
 \setcounter{equation}{0}
\renewcommand{\theequation}{9.\arabic{equation}}

The higher operators $d_r$ for $r=3, 4,5,...$ vanish due to
(\ref{E31}), (\ref{papro}) and (\ref{lam}) (these operators are written down
explicitly in terms of $\d_{0}$ and $\d_{2}$ in \cite{cmp1}).  Hence, the
spectral sequence collapses at this point, resulting in $ E_{\infty} = E_3 $,
and the defining equations of $E_{\infty}$ are:
\be
\L_{\a} E_{\infty} = 0 \la{lanew}
\ee
\be
( \pa_{\m})^{\dag} E_{\infty} = 0 \la{dagnew}
\ee
\be
n N E_{\infty} = 0  \la{chiralnew} \ee
\be
\left( N+ {\bar N} \right) \e_{\mu} E_{\infty}=0 \la{enew}
\ee
\be
\P_2 {\na}_{\dot \b} E_{\infty} = 0 \la{nabnew}
 \ee
\be
{\ov n}  [ 4 N(F) + 2 N(\y) - 4] E_{\infty} = K E_{\infty} = 0
\la{E33new}
\ee
and their complex conjugates. In addition, for $\ov n\neq 0$ ($n\neq0$) the
objects in $E_{\infty}$ have to be symmetric in their dotted (undotted)
indices, as discussed above.

We can construct solutions of these equations using superfields
in the  following way.
The basic chiral superfield is:
\be
  S(x, \q, {\ov \q}) = A(y)+\q \y(y) +\frac{1}{2}\q^2 F(y)
\la{sssufi}
\ee
where
\be
y^{\m} = x^{\m} + \frac{1}{2} \q^{\a} \s^{\m}_{\a \dot \b} {\ov \q}^{\dot \b}.
\ee
To find a solution of the above equations for $E_{\infty}$, we simply take any
polynomial $P(S,\pa,{\ov c})$ of the superfields, the derivative operator and
the antighost $\ov c$. Then we choose its `F' component, i.e. the coefficient
of $\q^2$ in the polynomial. Let us denote this $f= [P(S,\pa,{\ov c})]_F$. Then
we claim that the following is a  solution of the above equations plus the
symmetry requirements discussed in the previous section:
\be
e = {\rm{Sym_{dotted \; indices}}} \;\;
\Pi_{\pa^{\dag}=0} f \Q \in E_{\infty}.
\la{e}
\ee
We believe that all solutions of these equations are obtained in this manner,
but will not attempt to prove that here.  We have shown that there are an
infinite set of objects in the cohomology space, and the missing proof would
only establish that there are no more.  The more interesting question now is
whether any of the known ones correspond to anomalies.

Now we show that, in fact, (\ref{e}) is in the cohomology space $E_{\infty}$.
To see this, we notice that the grading operator
\be
N'_{\rm grad}=N_{\rm grad}+N(\q)
\ee
where $N_{\rm grad}$ is given by (\ref{gradop}),
satisfies
\be
N'_{\rm grad} S = 3 S,
\ee
so each term in (\ref{sssufi}) has
eigenvalue $N'_{\rm grad}=3$. Obviously, for a product of $k$ $S$ fields
(including arbitrary number of space-time derivatives),
$N'_{\rm grad}S^{k}=3kS^k$ holds. In terms of this new grading, operator $K$ in
(\ref{E33new}) becomes
\be
K = \ov n[2N(\q)+2N(c)+2(3N-N'_{\rm grad})-4].
\ee
Recalling  $N S^k=kS^{k}$, we
note that
\be
(3N-N'_{\rm grad}) P = 0.
\ee
Note that this would not be true in general if $P$ contained any
covariant derivatives $D_{\a}$.
Using this result we see that
\be
K P = \ov n[2N(\q) -4] P.
\ee
Now $P$ is not  homogeneous under the action of
$K$.  This equation shows that the terms of
$P$ which are homogeneous in $\q$ are
also homogeneous under the action of $K$.
In fact, the ${\q}^2$ term of P is homogeneous of degree
zero when acted upon by the operator $K$.
So we find that the operator $K$ acting on the the
${\q}^{2}$ term of any product of chiral superfields involving only partial
derivatives vanishes assuring that the $F$ component (with ${\q}^2$) of this
product is a solution of (\ref{E33new}).

Let us generalize our analysis to the case where
there are several superfields
$S_a$.  Then a more explicit form of our solution is:
\be
e = {\rm{Sym_{dotted \; indices}}}\int d^{2} \theta \,
\P_{\pa^{\dag} =0}     {\rm{P}}\left(S_{a_1},\pa_{ \a \dot \a} S_{a_2},
  \pa_{ \b \dot \b} \pa_{ \g \dot \g} S_{a_3},...\right)
                 \ov c_{\dot \k_{1}} \, \ov c_{\dot \k_{2}} ... ,
\ee
and the undotted indices are `free'. In above expression, $\pa_{\a \dot \b}=
\pa_{\mu}\s^{\mu}_{\a \dot \b}$ and P represents any  polynomial of its
argument
fields. Acting with $\Pi_{\pa^{\dag}=0}$ on any expression
simply corresponds to a
subtraction of the total derivative part. $e$,
as defined above, satisfies
(\ref{dagnew}),  (\ref{chiralnew}), (\ref{enew}) and  (\ref{E33new})
trivially; (\ref{lanew}) and (\ref{nabnew}) need to be checked explicitly.
As stated  above, any product of $S$'s is homogenous in $N'_{\rm grad}$,
therefore, its $F$ component is also homogeneous in $N'_{\rm grading}$ and
$N_{\rm grading}$, leading to $f=f_k$ with $N_{\rm grading}f_k=kf_k$.
As a consequence, the supersymmetry transformation $\d'=\d -\e^{\mu}\pa_{\mu}
 +c^{\a}\s^{\mu}_{\a {\dot \b}} \ov c ^{\dot \b}(\e^{\mu})^{\dag}=
c\L +\ov c \ov \L
 +c \ov \nabla +\ov c \nabla$, acting on $f_k$ splits into two independent
parts:
\be \d' f_k=(\d'_0+\d'_2)f_k=\pa_{\m}X^{\m}_k+\pa_{\mu}X^{\mu}_{k+2},
\la{sutra}
\ee
where $\d'_0 = c\L +\ov c \ov \L$ and $\d'_2 = c \ov \nabla +\ov c \nabla $;
$X^{\mu}_{k}$ and $X^{\mu}_{k+2}$ don't need to be further specified.
(\ref{sutra}) reiterates the well-known fact that the $F$ component of any
superfield transforms into a total derivative.
Using (\ref{lapa}) and (\ref{ladagpa}), we find that
$[\L,\Pi_{\pa^{\dag}=0}]=0$ and, therefore,
\be
\L_{\a} e=\L_{\a}\Pi_{\pa^{\dag}=0}f_k \Q=\Pi_{\pa^{\dag}=0}\L_{\a}
   f_k \Q
   =\Pi_{\pa^{\dag}=0}\pa_{\mu}X_{\a,k-1}^{\mu} \Q=0 ,
\ee
since $\Pi_{\pa^{\dag}=0}\pa=0$. Hence, (\ref{lanew}) is fulfilled.
Using (\ref{pinapi}), we see (\ref{nabnew}) is obeyed as well:
\be
\Pi_{2}\nabla_{\dot \b}e =
\Pi_{2}\nabla_{\dot \b}\Pi_{2}\Pi_{\pa^{\dag}=0}f_k \Q=
\Pi_{2}\nabla_{\dot \b}\Pi_{2}f_k \Q=
\Pi_{2}\nabla_{\dot \b} f_{k} \Q=
\Pi_2 \pa_{\mu}X_{\dot \b,k+1}^{\mu} \Q
  =0.
\ee
This proves that the objects of the form (\ref{e})
(or their complex conjugates)
are indeed in $E_{\infty}$.

The states with $n=\ov n =0$ can be built with chiral and antichiral
superfields, supercovariant derivatives $D_{a}$ and $\ov D_{\dot \a}$, and
their anticommutator $\pa_{\a \dot \b}$. This result reflects the fact that
the usual supersymmetric actions with ghost charge zero are BRS-invariant.

The one-to-one relation between $E_{\infty}$ and $H$ \cite{cmp1} leads to
the following objects ${\cal {X} }$ in the cohomology space:
\bea
{\cal {X} } & = &  \int d^{4}x f \nonumber \\
     & = & {\rm{Sym_{dot. \; ind.}}}
        \int d^4x \, \int d^{2} \theta \,
          {\rm{P}}\left(S_{a_1},\pa_{ \a \dot \a} S_{a_2},
  \pa_{ \b \dot \b} \pa_{ \g \dot \g} S_{a_3},...\right)
                 \ov c_{\dot \k_{1}} ... \, .
\la{finsol}
\eea
(The operator $\Pi_{\pa^{\dag}=0} \Q$ has been mapped into $\int d^{4}x$.)

\section{Conclusion}
\setcounter{equation}{0}
\renewcommand{\theequation}{10.\arabic{equation}}

We have shown that there are no polynomials in $H$ for the Wess-Zumino theory
that contain both $c$
and $\ov c$. The complex conjugate of every solution of the defining equations
for $E_{\infty}$ is also a solution. Hence, we can restrict the discussion of
the complete cohomology space to objects containing only the antighosts $\ov
c$.
Revealing the index structure explicitly, we have shown that
there is an infinite set of states of the form
\bea
{\cal {X} } _{\a_{1}...\a_{k_n}\dot \b_1...\dot \b_{k_n+g}} & = &
\mbox{Sym}_{\dot \b_{1}...\dot \b_{k_n+g}} \int d^{4}x \, d^{2} \q
\Bl  \{\pa_{{\a}_1 \dot\b_1 }...\pa_{{\a}_{k_1} \dot{\b}_{k_1} }
 S_{a_1}... \nonumber \\
& & ...
\pa_{{\a}_{k_{(n-1)}+1} \dot\b_{k_{(n-1)}+1} }...
\pa_{{\a}_{k_n} \dot\b_{k_n} } S_{a_n} {\ov c}_{\dot \b_{k_n+1}}
{\ov c}_{\dot \b_{k_n+2}}...\ov c_{\dot \b_{k_n+g}}
\Br \}.
\eea
in the cohomology space of the Wess-Zumino model.
The corresponding complex conjugate expressions are obtained by converting
dotted into undotted indices and vice versa, and $S \ra \ov S, \; \ov c \ra c$.
By contraction and symmetrization of the undotted indices, we can decompose
$ {\cal{X} }$ into operators of the form
\be
{\cal {A}}_{(a,b)}={\cal {A}}_{(\a_1 \a_2 \cdots \a_{a}) ({\dot \b}_{1}
{\dot \b}_{2} \cdots
{\dot \b}_{b}) },
\la{irrep}
\ee
where $b=k_n+g$ and $k_n-a$ is even and greater than or equal to $0$, since
contractions always involve pairs of undotted indices. In particular, we are
interested in polynomials with ghost charge $g=1$, which correspond to
anomalies. For these objects, we find $b-a$ is  odd and positive. Therefore,
the operators ${\cal {A}}$  are spinors.

These objects could appear as anomalies in the renormalization of composite
operators with the same spin structure as the anomaly. To compute the anomalies
of a given such composite operator, a term
of the form
\be  S_{ \Y}=
\int \, d^{4}x \, d^{4}\q \, \left[ {
  \Y_{\a_{1}...\a_{a}\dot\b_1...\dot\b_{b}}}
{ \F^{\a_{1}...\a_{a}\dot \b_1...\dot \b_{b}}} \right]
\label{source}
\ee
would be introduced into the action.
Here $ \Y$ is a composite operator with ghost charge zero and
${\F^{\a_{1}...\a_{a}\dot \b_1...\dot \b_{b}}}$ is a chiral source superfield.
There is a matching between the indices of the anomaly and those
of the anomalous operator, because both must couple to the
source $\F$.
Now to compute the anomaly, some specific form for the
composite operator $ \Y$ would be chosen
and then the one particle irreducible
generating functional $\G$ including one vertex (\ref {source}) would
be calculated. If there is an anomaly, one would find that
the supersymmetric variation of this part of $\G$
would be of the form
\be
 \d \G= \k \int \, d^{4}x \, d^{2}\q \, \left[ {
  {\cal {A} }_{\a_{1}...\a_{a}\dot\b_1...\dot\b_{b}}}
\F^{\a_{1}...\a_{a}\dot \b_1...\dot \b_{b}} \right]
\la{ggg}
\ee
where $\k$ is a calculable coefficient.

Note that (\ref{ggg}) is still in the cohomology space because $ \F$
transforms simply as a chiral superfield as does $S$. Hence, our derivation
of the cohomology applies to $\F$ just as it did to $S$. The indices of $ \F$
play no role in the cohomology discussion.

Because all possible anomalies have half-integer spin (see discussion of
(\ref{irrep})), it follows that all operators which can be anomalous also
have half-integer spin. Generally, the entire class of spinor operators in
supersymmetric theories containing chiral matter can be anomalous.

It would be interesting to extend the results of the present paper to the BRS
cohomology of the supersymmetric Yang-Mills theory.  This appears
to be rather difficult, because the gauge symmetry mixes with
supersymmetry in a tricky way. However, it is important.  In fact,
we conjecture that if there
are one-loop supersymmetry anomalies in rigid supersymmetric
theories, they will require at least one gauge propagator in the diagram--
i.e. they will occur only in supersymmetric
theories where chiral matter is coupled to gauge theory.

\vfill
\ed